\documentclass[aps,prc,10pt,twocolumn,superscriptaddress,preprintnumbers,amsmath,amssymb]{revtex4-1}
\usepackage[dvips]{graphicx}
\usepackage{dcolumn}
\usepackage{bm}
\usepackage{color}
\usepackage{hyperref}

\begin{document}
\title{A formally exact one-frequency-only Bethe-Salpeter-like equation.\\
Similarities and differences between $GW$+BSE and self-consistent RPA}
     
\author{Valerio Olevano}\email{valerio.olevano@neel.cnrs.fr}
\affiliation{Universit\'e Grenoble Alpes, 38000 Grenoble, France}
\affiliation{CNRS, Institut N\'eel, 38042 Grenoble, France}
\affiliation{European Theoretical Spectroscopy Facility (ETSF)}

\author{Julien Toulouse}\email{toulouse@lct.jussieu.fr}
\affiliation{Laboratoire de Chimie Th\'eorique (LCT), Sorbonne Universit\'e and CNRS, F-75005 Paris, France}

\author{ Peter Schuck}\email{schuck@ipno.in2p3.fr}
\affiliation{Institut de Physique Nucl\'eaire, IN2P3-CNRS,
Universit\'e Paris-Sud, 91406 Orsay, France}
\affiliation{Universit\'e Grenoble Alpes, 38000 Grenoble, France}
\affiliation{CNRS, Laboratoire de Physique et de Mod\'elisation des Milieux Condens\'es, 38042 Grenoble, France}

\date{February 4, 2019}

\begin{abstract}
A formally exact Bethe-Salpeter-like equation for the linear-response function is introduced with a kernel which depends only on the one frequency of the applied field. This is in contrast with the standard Bethe-Salpeter equation (BSE) which involves multiple-frequency integrals over the kernel and response functions.
From the one-frequency kernel, known approximations are straightforwardly recovered.
However, the present formalism lends itself to more powerful approximations.
This is demonstrated with the exact analytical solution of the Hubbard molecule.
Similarities and differences of the $GW$+BSE approach with the self-consistent random-phase approximation (RPA) is also discussed.
\end{abstract}

\maketitle

\section{Introduction}

The development of efficient many-body approaches is an active research field in quantum chemistry and various branches of physics, such as condensed-matter, cold-atom, atomic, molecular, and nuclear physics. Originally developed in the framework of subnuclear and nuclear physics to describe bound states of systems of two interacting particles like the deuteron, the Bethe-Salpeter equation (BSE) \cite{SalpeterBethe51} has become an approach commonly used also in solid-state and condensed-matter physics \cite{HankeSham74,HankeSham75,HankeSham79,Strinati82,OnidaAndreoni95,AlbrechtOnida98,RohlfingLouie98,BenedictShirleyBohn98,OlevanoReining01}, atomic physics \cite{LiOlevano17}, and quantum chemistry \cite{JacqueminBlase15,BrunevalNeaton15,BaumeierRohlfing12}.

The fact that the standard BSE can be demonstrated \cite{CasalbuoniOlevano10} to be equivalent to the Ward identities and the Hedin integral equation for the vertex \cite{Hedin} enables a natural transfer of approximations, i.e.\ the Hedin $GW$ approximation \cite{Hedin} on the self-energy toward the BSE kernel.
The idea behind the $GW$ approach to tackle correlations simply by the introduction of only screening, i.e.\ the simple replacement of the bare two-body interaction $v$ by a screened interaction $W$, can be directly transferred to an approximation to the irreducible BSE kernel, which is hence written as $W$ instead of the time-dependent Hartree-Fock (TDHF) exchange kernel.
In contrast with the TDHF exchange kernel, which for electronic systems is the opposite of the static Coulomb interaction $v(r, r')=1/|r- r'|$, a BSE kernel at the same level of the $GW$ approximation should in principle be frequency dependent since it relies on the dynamically screened Coulomb interaction $W(r,r',\omega)$.
This frequency dependence, which can be worked out, though with some difficulties, when calculating the $GW$ self-energy, and which is an important ingredient to have quasiparticle energies more in agreement with experiment, implies multiple-frequency integrals in the BSE and represented so far an insurmountable obstacle to the resolution of the full BSE.
For this reason almost all BSE calculations were obliged to neglect the dynamical dependence of the BSE kernel and solve a static BSE.
This approach often called $GW$+BSE \cite{MRC} which uses a dynamical $W(r,r',\omega)$ in the self-energy and a static $W(r,r',\omega=0)$ in the BSE kernel has nevertheless provided good results in agreement with experiment and exact solutions \cite{OnidaAndreoni95,AlbrechtOnida98,RohlfingLouie98,BenedictShirleyBohn98,OlevanoReining01,LiOlevano17}.

It is difficult to estimate how important can be dynamical effects beyond the static BSE.
Nevertheless, it is often conjectured that deviations of the static BSE solution from experiment can be solely due to dynamical BSE effects.
A tentative list might include effects associated to double excitations in quantum chemistry \cite{RebTouSav-INC-13} or to electron-hole screened interaction in metals \cite{MariniDelSole03}.
Efforts to study dynamical BSE effects and introduce a real frequency dependence into the BSE have recently been attempted \cite{Pina,Pina2,ZhaSteYan-JCP-13,Julien}.
The standard BSE is an equation over two-body Green/correlation functions (kernel and response functions), i.e.\ functions of four space-time points.
In systems with time-translation invariance, there is one-time degree-of-freedom less, which means always functions of three time differences, or their three Fourier transformed frequencies.
The full dynamical BSE involves a so far numerically intractable integration over frequencies in the kernel and in the response function.
Recent efforts \cite{Pina,Pina2} have tried to redefine a kernel which incorporates frequency integration, to finally arrive at a more easily solvable one-frequency equation.
Another approach \cite{ZhaSteYan-JCP-13,Julien} has considered the coupling of the linear-response function to uncorrelated two-particle-two-hole (\textit{2p-2h}) states.
The coupling of the linear-response function to collective states plus free particle-hole (\textit{p-h}) states to account for double excitations has been discussed in Ref. \cite{Pina}.

All previous works followed the route which starts from the multi-frequency standard BSE and tries to reduce the number of involved frequencies, so as to end up with an equation with just only the one frequency of the external field.
The purpose of this work is to follow a different route: we introduce \textit{from the beginning} a formally exact one-frequency BSE-like equation, i.e.\ depending only on the frequency of the external field, for a linear-response function. In particular, this means that also the integral kernel $K$ depends only on the one frequency of the external field. Explicit expressions for $K$ will be elaborated in terms of well defined correlation functions and higher Green functions. For readers interested right away to see the final result, they may consult Eqs.~(\ref{response}), (\ref{kernel2-total}), (\ref{kernel2-static}), (\ref{kernel2}), and (\ref{BSE}). Starting from these expressions  we
then rederive the approximate expressions given in the literature mentioned above.
However, since our expressions are more general, they lend themselves to more far-reaching approximations without loosing the advantage of a one-frequency only approach.
This is demonstrated with the exact solution of the Hubbard molecule.
But we will also point out that the response function calculated in this way keeps all desirable qualities of the standard random-phase approximation (RPA), such as fulfillment of sum rules and conservation laws.
Our derivation is based on  the equation-of-motion (EOM) technique applied to an appropriately defined four-point one-frequency linear-response function.

The paper is organized as follows. We will use the example of the EOM technique for establishing the Dyson equation for the one-body Green function presented in Sec.~\ref{spdyson} to introduce the key points of the derivation of the one-frequency-only BSE-like equation which will then be presented in Sec.~\ref{onefreqbse}.
In Sec.~\ref{comparison} we will establish the connection of the present formalism to the previous approaches of Refs.~\cite{Pina,Pina2,Julien} and with the standard $GW$+BSE approximation, making parallels also with the self-consistent random-phase approximation (SCRPA)~\cite{NPA512, Hubbard, 3-Lipkin, Tohy1}. We will present in Sec.~\ref{Hubbardmol} a short application of our formalism to the Hubbard molecule which in this way can be solved exactly. 
Finally, Sec.~\ref{conclusions} contains our conclusions and outlook. 

Atomic units are used throughout this work.

\section{Rederivation of the one-body Dyson equation}
\label{spdyson}

To set the stage, we first present a short derivation of the Dyson equation for the standard one-body Green function by the EOM technique, highlighting the points over which we will base the derivation of the one-frequency-only BSE-like equation in the next section.

We consider the most generic Hamiltonian, $H = H_0 + V$, composed by the non-interacting (kinetic plus external potential) Hamiltonian $H_0$ and the two-body interaction operator $V$, which we write in terms of creation/annihilation operators $c^\dag_k$ and $c_k$ on an arbitrary orthonormal basis set of orbitals $\{\phi_k(r)\}$ as
\begin{eqnarray}
  H = \sum_{k_1 k_2} \epsilon_{k_1 k_2} c^\dag_{k_1} c_{k_2} + \frac{1}{4}\sum_{k_1k_2k_3k_4}
      \bar v_{k_1k_2k_3k_4}c^\dag_{k_1}c^\dag_{k_2}c_{k_4}c_{k_3},
\nonumber\\
\label{H}
\end{eqnarray}
where $\epsilon_{k_1 k_2}$ are the matrix elements of the non-interacting Hamiltonian $H_0$ over the orthonormal basis set, and 
\begin{equation}
  \bar{v}_{k_1k_2k_3k_4} = \langle k_1k_2|v|k_3k_4\rangle - \langle k_1k_2|v|k_4k_3\rangle,
  \label{as-v}
\end{equation}
are the antisymmetrized matrix elements of the Coulomb interaction $v(r,r') = 1/|r- r'|$, or more precisely, detailing the notation,
\begin{eqnarray}
 \langle k_1k_2|v|k_3k_4\rangle = \! \int \!\! d r dr' \, \phi^*_{k_1}(r) \phi^*_{k_2}(r') v(r, r') \phi_{k_3}(r) \phi_{k_4}(r'). 
\nonumber\\
\end{eqnarray}
We work in a three-dimensional space and $r$ and $k$ are meant as 3D vectors, but we can generalize to 1D and 2D;
the spin degree of freedom $\sigma$ is always implied and can be included for spin-polarized cases in $k$ and in $r$ and summed over whenever $r$ is integrated out.

We remind the definition of the one-body Green function
\begin{eqnarray}
G_{kk'}(t-t') = -i \langle 0 | \mathrm{T}\{ c_k(t) c^\dag_{k'}(t') \} | 0 \rangle,
\end{eqnarray}
where $\mathrm{T}\{o(t) o'(t')\} = \theta(t-t') o(t) o'(t') - \theta(t'-t) o'(t') o(t) $ is the time-ordering product between fermion operators $o$, $c_k(t) = e^{iHt}c_ke^{-iHt}$ is the time-dependent annihilation operator in the Heisenberg formalism (and similarly for the time-dependent creation operator $c^\dag_{k'}(t')$), and $|0 \rangle$ is the ground state.
We can then introduce the non-interacting Green function $G^0$, associated to the non-interacting Hamiltonian $H^0$, and its inverse
\begin{eqnarray}
  G^{0^{-1}}_{kk'}(t-t') = \delta(t-t') (\delta_{kk'} i \partial_t - \epsilon_{kk'})\,,
\end{eqnarray}
by which we can write out a first EOM for $G$
\begin{eqnarray}
  \sum_{k_1} \int dt_1 &\,& G^{0^{-1}}_{kk_1}(t-t_1) G_{k_1k'}(t_1-t') = \nonumber \\
  && \delta_{kk'}\delta(t-t') -i\langle 0|\mathrm{T}\{j_k(t)c^\dag_{k'}(t')\}|0\rangle,
\label{1st-eom} 
\end{eqnarray}
where we have introduced the operator
\begin{equation}
j_k = 
[c_k,V] = \frac{1}{2}\sum_{k_2k_3k_4}\bar v_{kk_2k_3k_4}c^\dag_{k_2}c_{k_4}c_{k_3}\,.
\label{j-k}
\end{equation}
The term containing $j_k$ is a two-body Green function with a particular time ordering.

For simplicity and without loss of generality, we will henceforth write the equations for the case of homogeneous systems, where $k$ stands for momentum (and spin) and $\epsilon_k$ is the kinetic energy.
This is very similar to work in the natural spin-orbital basis, also sometimes called canonical basis, obtained from the diagonalization of the one-body density matrix in the case of inhomogeneous or finite systems.
Let us now write the well-known Dyson equation \cite{FW}
\begin{equation}
(i\partial_t - \epsilon_k) G_k(t-t') = \delta(t-t') +\int dt_1 \Sigma_k({t-t_1})G_k({t_1-t'}).
\label{Dyson}
\end{equation}
Using Eq.~(\ref{1st-eom}), the self-energy is then formally given by
\begin{equation}
\Sigma_k({t-t'}) =  -i\int dt_1\langle 0|\mathrm{T}\{ j_k(t)c^\dag_k(t_1) \} |0\rangle G^{-1}_k(t_1-t'),
\label{self}
\end{equation}
where we introduced the inverse of the Green function defined  by $G^{-1}G = 1$ in short-hand notation.
From the Dyson equation (\ref{Dyson}), this inverse can be expressed as
\begin{equation}
  G^{-1}_k(t-t') =  G^{0^{-1}}_{k}(t-t') - \Sigma_k({t-t'}).
  \label{GF-inv}
  \end{equation}
The self-energy can therefore be written as
\begin{eqnarray} 
\Sigma_k({t-t'}) &=& -i\int dt_1\langle 0|\mathrm{T}\{ j_k(t)c^\dag_k(t_1) \}|0\rangle
\nonumber\\
&&\;\;\;\;\times\left[G_k^{0^{-1}}(t_1-t') - \Sigma_k({t_1-t'})\right].
 \label{self2}
  \end{eqnarray}
In this equation, $G_k^{0^{-1}}(t_1-t')$ can be applied on the left using a second EOM (we should realize that $\partial_{t_1}$ contained in $G_k^{0^{-1}}(t_1-t')$ normally acts to the right and, thus, one has to perform an integration par parts over $t_1$ to make it act to the left, which changes $i\partial_{t_1}$ into $-i\partial_{t_1}$), we then arrive at
\begin{equation}
  \Sigma_k({t-t'}) = T_k({t-t'}) -  C^\mathrm{red}_k({t-t'}),
\label{self3}
\end{equation}
where
\begin{equation} T_k({t-t'}) = V_k^\mathrm{MF}\delta(t-t') -i\langle 0|\mathrm{T} \{ j_k(t)j_k^\dag(t') \} |0\rangle,
  \label{T1}
\end{equation}
which is a kind of \textit{one-body T-matrix}, and
\begin{eqnarray}
  C^\mathrm{red}_k(t-t')&=& (-i)^2 \int dt_1 dt'_1 \, \langle 0|\mathrm{T} \{ j_k(t)c_k^\dag(t_1) \} |0\rangle \nonumber\\ 
  &&\times G^{-1}_k(t_1-t'_1)\langle 0|\mathrm{T} \{ c_k(t'_1)j_k^\dag(t') \} |0\rangle. \;\;
  \label{C-red}
  \end{eqnarray}
The usual mean-field potential is given by
\begin{equation}
  V_k^{{\rm MF}} =\langle 0|\{[c_k,V],c^\dag_k\}|0\rangle = \sum_{k'}\bar v_{kk'kk'}n_{k'},
  \label{MF}
  \end{equation}
where $\{..\}$ stands for the anticommutator and 
\begin{equation}
  n_k = \langle 0|c^\dag_kc_k|0\rangle
  \label{occs}
  \end{equation}
are the occupation numbers. We should mention that the mean-field potential in Eq. (\ref{MF}) is only diagonal in a homogeneous system. In a finite system, this is not necessarily the case in spite of the fact that in the natural spin-orbital (canonical) basis the density matrix is diagonal. However, to avoid heavy formulas, we will always from now on assume that the mean-field is also diagonal.
It is easily recognized that the second term of the expression for the above one-body $T$-matrix is expressed as a three-body propagator of the two-particle-one-hole (2p-1h) plus two-hole-one-particle (2h-1p) type.
This 3-body Green function contains the so-called one-line reducible Feynman graphs, which is easily verified by perturbation theory.
By definition, a self-energy should not contain such contributions which can be ``cut'' into two pieces by cutting a single fermion line at a given time.
It is again easily verified by perturbation theory that the second term on the right-hand side in Eq. (\ref{self3}) just does nothing else than taking out of the $T$-matrix all reducible terms.
Therefore, in short, we can write the self-energy as
\begin{equation}
 \Sigma_k(t-t') = V_k^{{\rm MF}}\delta(t-t') -i\langle 0|\mathrm{T} \{ j_k(t)j_k^\dag(t') \}
|0\rangle^\mathrm{irr}, 
\label{self4}
\end{equation}
where the index ``irr'' indicates that the corresponding correlation function should be one-line irreducible.
Expression (\ref{self4}) is therefore a formally exact and compact expression for the self-energy.
Please also note that the expression is very symmetric, which is well suited for introducing approximate forms of the self-energy.
For completeness, let us also write an expression for the one-body Green function in the following way (with $G^{0^{-1}}G^{0} = 1$)
\begin{equation}
 G_k = G_k^{0} + G_k^{0}T_kG_k^{0}~~{\rm with}~~~ T_k = \Sigma_k +\Sigma_kG^0_kT_k \,.
\label{Lipp}
\end{equation}
Notice that in Eq.~(\ref{self4}) we have an index ``irr'', so that the single particle $T$-matrix of Eq.~(\ref{Lipp}) is different from $\Sigma$.
Here, we did not write out the time dependencies and integrals.
In frequency space there are no integrations and it becomes an algebraic equation as, by the way, the Dyson equation itself.
Please note that Eq.~(\ref{Lipp}) has the usual form connecting a Green function to the scattering $T$-matrix.
However, here the $T$-matrix is defined for a many-body system.
Taking out of $T$ the one-line reducible contributions changes $TG^{0}$ into $\Sigma G$, that is we also have the relation $T=\Sigma + \Sigma G^0T$ as indicated in Eq.~(\ref{Lipp}).

After this hopefully pedagogic and relatively elaborate presentation of well-known many-body relations on the one-body Green function, let us now turn to the two-body case and response function.

\section{Response function and Bethe-Salpeter-like equation} 
\label{onefreqbse}

\subsection{Derivation of the one-frequency Bethe-Salpeter-like equation}

We will derive for the two-time response function defined by, with $k_1 \ne k_2$ and $k_1' \ne k_2'$,
\begin{equation}
 R_{k_1k_2k'_1k'_2}(t-t') =  -i \langle 0| \mathrm{T}\{ c^\dag_{k_2}(t)c_{k_1}(t)
c^\dag_{k'_1}(t')c_{k_2'}(t') \} |0\rangle 
\label{response}
\end{equation}
an \textit{exact} equation which has the same structure as the Dyson equation for the one-body Green function derived above. The inequalities $k_1 \ne k_2$ and $k_1' \ne k_2'$ are not independent of the one-body basis: for homogeneous matter the indices stand for momenta and spin and then the inequalities concern the momenta. For finite systems the indices correspond to the canonical basis. With this definition, we have $\langle 0|c^{\dag}_{k_2}c_{k_1}|0\rangle =0$ and $\langle 0|c^{\dag}_{k_1'}c_{k_2'}|0\rangle =0$, so that the quantity $R$ is the same as the linear-response function often denoted by $\chi$ in condensed-matter physics or quantum chemistry~\cite{RebTouSav-INC-13}.
Let us further note that in Eq.~(\ref{response}) we have chosen a definite ordering of the fermion operators.
This stems from the fact that we are considering a one-body-density-matrix/one-body-density-matrix correlation function.
Notably there will appear an integral kernel which also depends on only one time difference or on one frequency. For people used to multi-time Green functions, this may seem surprising. However, this is not unknown in the literature \cite{NPA628}. There also exists, e.g., the Mori-Zwanzig formalism for correlation functions  of statistical physics \cite{Mori, Zwanzig}. Furthermore, in nuclear physics, the EOM formalism developed by Rowe \cite{Rowe}, and further developed in Refs.~\cite{NPA628,NPA512,Hubbard,3-Lipkin,Tohy1} (with more references therein), is closely related to what we will present here. However, these facts seem to be very little known in the condensed-matter and chemical physics communities where one often struggles to get rid of eventually superfluous frequency dependencies of the integral kernel of the response function which are inherent to the so-called Hedin equations \cite{Hedin}. Introducing a single frequency integral kernel from the start and not a posteriori will turn out to have several advantages. For example, though we will recover, e.g., certain aspects of the $W$ kernel of the BSE as used in the $GW$ approach, we will also see more clearly what kind of approximations are involved with the use of static and dynamic forms of $W$ in the BSE and how eventually to go beyond in a systematic way.

So, let us start as before by writing down the first EOM for the response function 
\begin{eqnarray}
  \int && dt_1 \, {\tilde R}^{0^{-1}}_{k_1k_2}(t-t_1) R_{k_1k_2k'_1k'_2}(t_1-t') = 
   N_{k_1k_2k_1'k_2'}\delta(t-t')
   \nonumber\\
   &&\:\;\;\;\; -i\langle 0|\mathrm{T} \{ J_{k_1k_2}(t)c^\dag_{k'_1}(t')c_{k'_2}(t') \} |0\rangle,
   \label{eom-1}
\end{eqnarray}
where 
\begin{equation}
  {\tilde R}^{0^{-1}}_{k_1k_2}(t-t') = \delta(t-t')( i \partial_t -\epsilon_{k_1} + \epsilon_{k_2}),
  \label{R0-inv}
\end{equation}
which is a straightforward extension of the one-body case. We have also introduced
\begin{eqnarray}
  J_{k_1k_2} &=&[c^{\dag}_{k_2}c_{k_1},V]
 \nonumber\\
  &=&  \frac{1}{2}\sum_{k'_2k'_3k'_4}\bar v_{k_1k'_2k'_3k'_4}c^\dag_{k_2}c^\dag_{k'_2}c_{k'_4}c_{k'_3}
    \nonumber\\
    & &
  +\frac{1}{2}\sum_{k'_1k'_2k'_3}\bar v_{k'_1k'_2k'_3k_2}c^\dag_{k'_1}c^\dag_{k'_2}c_{k'_3}c_{k_1}
    ,
  \label{J-kk'}
\end{eqnarray}
and the so-called norm kernel 
\begin{eqnarray}
  N_{k_1k_2k_1'k_2'} &=& \langle 0 | [c^\dag_{k_2}c_{k_1},c^\dag_{k'_1}c_{k'_2}] | 0 \rangle
  \nonumber \\ &=&
  \delta_{k_1k'_1}\delta_{k_2k'_2}N_{k_1k_2},
\label{norm}
\end{eqnarray}
with
\begin{equation}
  N_{k_1k_2} = n_{k_2} - n_{k_1}~ = |n_{k_2} - n_{k_1}|N^0_{k_1k_2},
  \label{norm2}
\end{equation}
where the sign factor $N^0$ is given by
\begin{equation}
 N^0_{k_1k_2} = 1~{\rm for}~ k_1 > k_2~{\rm and}~-1~{\rm for}~k_1 < k_2,
\label{F-0}
\end{equation}
and therefore $N^0_{k_1k_2}N^0_{k_1k_2}=1$. 
Please note that the one-body density matrix $\langle 0|c^\dagger_{k_2}c_{k_1}|0\rangle$ is diagonal for our assumed homogeneous system (or in the canonical basis) and we suppose that it is also diagonal in spin.
One recognizes in Eqs.~(\ref{norm})-(\ref{norm2}) the phase-space factors from the standard RPA when the occupation numbers $n_k$ are replaced by their step function form, $n^0_k$, when using the Hartree-Fock (HF) ground state. In general, however, the occupation numbers are the correlated ones, different from zero and one. It is remarked that this norm factor is a different feature with respect to the one-body Green-function case. Note also that, contrary to the one-body case, the quantity ${\tilde R}^{0^{-1}}$ introduced in Eq.~(\ref{R0-inv}) is not exactly the inverse of the non-interacting response function $R^0$, but instead we have in short-hand notation ${\tilde R}^{0^{-1}} R^{0} = N$ where $N$ is the norm matrix.

We now proceed exactly in analogy with the one-body case. Because of the presence of the norm matrix $N$ in Eq. (\ref{eom-1}), we first have to divide it out by multiplying Eq. (\ref{eom-1}) by the inverse of $N$. Writing Eq. (\ref{eom-1}) schematically as
\begin{eqnarray}
{\tilde R}^{0^{-1}}R = N + F,
\end{eqnarray}
we obtain by division with $N$
\begin{eqnarray}
{\tilde R}^{0^{-1}}\tilde R = 1 + \tilde F~=~ 1+\tilde F\tilde R^{-1}\tilde R~\equiv 1 + K\tilde R,
\end{eqnarray}
with $\tilde R = RN^{-1}$ and $\tilde F = FN^{-1}$. So we arrive at a BSE-like equation of the form
\begin{eqnarray}
{\tilde R}^{-1} = {\tilde R}^{0^{-1}} - K,
\end{eqnarray}
with the kernel $K$ given by
\begin{eqnarray}
K = \tilde F \tilde R^{-1}~=~\tilde F[{\tilde R}^{0^{-1}} - K].
\end{eqnarray}
With explicit notations, the BSE-like equation with a one-frequency kernel can thus be written as
\begin{eqnarray}
  && \int dt_1 {\tilde R}^{0^{-1}}_{k_1k_2} (t-t_1) \tilde R_{k_1k_2k'_1k'_2}(t_1-t')  = \delta_{k_1k'_1}\delta_{k_2k'_2}\delta(t-t')
  \nonumber\\
  && + \int dt_1\sum_{k_3k_4} K_{k_1k_2k_3k_4}(t-t_1)\tilde R_{k_3k_4k'_1k'_2}(t_1-t'),
  \label{R-eom}
\end{eqnarray}
with
\begin{eqnarray}
     K_{k_1k_2k'_1k'_2}(t-t') = -i\int dt_1\sum_{k'_3k'_4} \;\;\;\;\;\;\;\;\;\;\;\;\;\;\;\;\;\;\;
  \nonumber\\  
    \langle 0|\mathrm{T} \{  J_{k_1k_2}(t)c^{\dag}_{k_3'}(t_1)c_{k_4'}(t_1) \} |0\rangle N^{-1}_{k_3'k_4'}
  \nonumber\\ 
    \;\;\; [{\tilde R}^{0^{-1}}_{k'_3k'_4}(t_1-t')\delta_{k'_3k'_1}\delta_{k'_4k'_2} -  K_{k'_3k'_4k'_1k'_2}(t_1-t') ].
  \label{kernel}
  \end{eqnarray}
We apply then the EOM a second time as in the one-body case and obtain the final expression of the kernel
\begin{eqnarray}
  K_{k_1k_2k'_1k'_2}(t-t') &=&K^0_{k_1k_2k'_1k'_2}\delta(t-t')
  +  K^\mathrm{dyn}_{k_1k_2k'_1k'_2}(t-t'),\nonumber\\
\label{kernel2-total}
\end{eqnarray}
with a purely static contribution
\begin{eqnarray}
  K^0_{k_1k_2k'_1k'_2}&=& \langle 0 | [[c^{\dag}_{k_2}c_{k_1},V],c^{\dag}_{k_1'}c_{k_2'}] | 0 \rangle N^{-1}_{k_1'k_2'}, \nonumber\\
\label{kernel2-static}
\end{eqnarray}
and a dynamic contribution
\begin{eqnarray}
    K^{\rm dyn}_{k_1k_2k'_1k'_2}(t-t')&=&
    -i\langle 0 | \mathrm{T} \{  J_{k_1k_2}(t)  J^\dag_{k'_1k'_2}(t')\} | 0 \rangle^\mathrm{irr}N^{-1}_{k_1'k_2'}.\nonumber\\
\label{kernel2}
\end{eqnarray}
Please note the complete analogy of this expression with Eq. (\ref{self4}). 
At this point some discussion is in order: we realize that the right-hand side of Eq. (\ref{kernel2}) corresponds to a four-body Green function of the \textit{2p-2h} and \textit{2h-2p} type. 
It contains therefore double \textit{p-h} excitations. The index ``irr'' may seem less evident than in the one-body case. 
One may, however, verify again by perturbation theory that everything works exactly as in the one-body case and that the subtraction of the matrix $K$ in ${\tilde R}^{0^{-1}} -  K$ exactly eliminates all \textit{p-h} reducible contributions of the \textit{2p-2h}/\textit{2h-2p} Green function. The \textit{p-h} (two-line) irreducibility is just the analog of the one-line irreducibility in the one-body case. Up to some technical details to be discussed below, we thus have derived, as announced, a BSE-like equation with a one-frequency kernel obtained by Fourier transforming into frequency space the time dependence of the kernel in Eq. (\ref{kernel2}), which at equilibrium depends only on the time difference $t-t'$. The frequency-space BSE-like equation that we have obtained is thus
\begin{eqnarray}
  &&
  ( \omega -\tilde \epsilon_{k_1} + \tilde \epsilon_{k_2}) \tilde R_{k_1k_2k'_1k'_2}(\omega) = \delta_{k_1k'_1}\delta_{k_2k'_2}
  \nonumber \\ &&
  + \sum_{k_3k_4} [K^0_{k_1k_2k_3k_4}  + K^\mathrm{dyn}_{k_1k_2k_3k_4} (\omega)]  \tilde R_{k_3k_4k'_1k'_2}(\omega),
\label{BSE-tildeR}
\end{eqnarray}
where
\begin{equation}
  \tilde{\epsilon}_k = \epsilon_k + V^{\rm MF}_k,
  \label{HF-e}
\end{equation}
are the one-body energies with mean-field shifts included. Please notice that in Eq. (\ref{BSE}) the kernel $K^0$ is now without the mean-field contribution, i.e. in Eq.~(\ref{kernel2-static}) $V$ has been replaced by $V - V^\text{MF}$ where $V^\text{MF}$ is the mean-field potential operator. Not to introduce new symbols, from now on, $K$ should always be understood in this way. 

This needs, however, further elaboration and  discussions. Actually the existence of the kernel $K$ hinges entirely on the existence of the inverse of the one-frequency response function $\tilde R$, via $ K = {\tilde R}^{{0}^{-1}} - \tilde R^{-1}$. Again, this is in complete analogy to the case of the Dyson equation: $\Sigma = G^{0^{-1}} - G^{-1}$. For readers who may doubt about the existence of $\tilde R^{-1}$, we announce that below we will find approximate expressions for $ K$ which reproduce known expressions from the literature which have been derived from the Hedin equations. We note that, as this was the case with the one-body self-energy, also here the single-frequency kernel in Eq. (\ref{kernel2-total}) splits into a purely static (instantaneous) and a dynamic (time-dependent) part. It is  quite suggestive to interpret the purely static term $ K^0$ as some kind of higher mean field. Below, we will give an explicit expression for it and will see that it contains static \textit{p-h} correlation functions. Viewing the ground state as containing a gas of \textit{p-h} quantum fluctuations, one can then interpret the purely static term as the (frequency-independent) mean field of those fluctuations. We will refer to $ K^0$ as a ``particle-hole mean field'' and shall show below in which way it is related to a specific form of $W$ in the $GW$+BSE approach. 

However, before that, let us transform the BSE-like equation by returning from $\tilde R$ to the original linear-response function $R$. It is straightforward to show that the latter then obeys the following equation
\begin{eqnarray}
  &&
  ( \omega -\tilde \epsilon_{k_1} + \tilde \epsilon_{k_2})R_{k_1k_2k'_1k'_2}(\omega) = N_{k_1k_2k'_1k'_2}
  \nonumber \\ &&
  + \sum_{k_3k_4} [K^0_{k_1k_2k_3k_4}  + K^\mathrm{dyn}_{k_1k_2k_3k_4} (\omega)] R_{k_3k_4k'_1k'_2}(\omega).
\label{BSE}
\end{eqnarray}

The reader may be worried that we  did not get rid of the possibly troublesome norm factor $N_{k_1k_2}=n_{k_2}-n_{k_1}$ in the denominator in Eqs.~(\ref{kernel2-static}) and (\ref{kernel2}) implying that there may be numerical troubles for situations where $n_{k_1}\simeq n_{k_2}$. Actually, there are good reasons for this division.
It is analogous to, e.g., what happens with the generator coordinate method (GCM) where also the norm kernel has to be diagonalized and configurations corresponding to zero eigenvalues be eliminated \cite{RS}. This always happens when expanding the quantity of interest into a non-orthogonal basis set (here the products $c^\dagger_{k_2}c_{k_1}$), a feature which is also underpinning our approach. Actually the present approach is practically equivalent to the EOM of Rowe \cite{Rowe} (see also Ref. \cite{Tohy1}), where one expands an excited state into a series of components where higher and higher many-body operators act on the formally exact ground state. Exactly the same type of norm factors $N$ as here appear on the right-hand-side of an eigenvalue problem as in Eq. (\ref{RPA-matrix}) below. We will not further discuss this very general case here as we will henceforth work in the \textit{p-h}/\textit{h-p} subspace (for definition, see below), where this problem does not appear, and replace the norm kernel in the denominator by its HF expression. Taking higher-order corrections of the norm in the denominator into account has probably little influence on the accuracy of the results as suggested by some explicit examples \cite{3-Lipkin}.

Before going on, let us transform our BSE-like equation into an eigenvalue problem. As just mentioned, this will be done in the \textit{p-h}/\textit{h-p} subspace
\begin{equation}
 \sum_{p'h'} \begin{pmatrix}A&B\\-B^*&-A^*
  \end{pmatrix}_{php'h'}
  \begin{pmatrix}X^{\nu}_{p'h'}\\Y^{\nu}_{p'h'}\end{pmatrix}
    =\Omega_{\nu}\begin{pmatrix}X^{\nu}_{ph}\\Y^{\nu}_{ph}\end{pmatrix},
    \label{RPA-matrix}
\end{equation}
with $h,h'$ referring to hole states ($h,h' \le k_{\rm F}$, where $k_{\rm F}$ is the Fermi momentum) and $p,p'$ referring to particle states ($p,p' > k_{\rm F}$).
This equation is of the typical RPA form as described, e.g., in Ref. \cite{RS}.
The present generic equation is, however, potentially much more general because in principle the $A$ and $B$ matrices depend on the eigenvalues $\Omega_{\nu}$ and on the amplitudes $(X^\nu, Y^\nu)$, the latters being related to the ground state $|0\rangle$ and excited state $|\nu\rangle$ by $X^{\nu}_{ph}= \langle 0|c^\dag_hc_p|\nu\rangle$ and $Y^{\nu}_{ph} = \langle 0|c^\dag_pc_h|\nu\rangle$. The $A$ and $B$ matrices are related to the one-frequency kernel $K$ in Eq. (\ref{kernel2-total}) by
\begin{eqnarray}
  A_{php'h'} &=& (\tilde \epsilon_p - \tilde \epsilon_h)\delta_{pp'}\delta_{hh'} +  K_{php'h'},\nonumber\\
  B_{php'h'} &=&  K_{phh'p'}.
  \label{A-B}
\end{eqnarray}
For example, to first order in the interaction this gives
\begin{equation}
   K_{php'h'} \rightarrow \bar v_{ph'hp'}~\text{and}~~  K_{phh'p'} \rightarrow \bar v_{pp'hh'},
  \label{K-1}
\end{equation}
where the occupation factors $n_i$ have been replaced by their uncorrelated form $n^0_i$, and Eq. (\ref{RPA-matrix}) reduces to the standard RPA equation (with exchange) or TDHF.
We will not further elaborate on the eigenvalue form of our approach and rather continue investigating the one-frequency kernel $K$.

\subsection{The purely static part of the kernel $K$}

Let us now discuss the $K^0$ term of the kernel and see how far it is related to the static $W$ kernel of the $GW$ approach. To establish an explicit form for $K^0$, we have to evaluate the double commutator contained in the particle-hole mean-field part of Eq. (\ref{kernel2-static}).
One finds
\begin{eqnarray}
  &&
    K^0_{k_1k_2k_3k_4} = N_{k_1k_2}\bar v_{k_1k_4k_2k_3}
  \nonumber\\ && \quad
  \Big[- \frac{1}{2}\sum_{ll'l''}(\delta_{k_2k_4} \bar v_{k_1ll'l''}C_{l'l''k_3l} +\delta_{k_1k_3}\bar v_{ll'k_2l''}C_{k_4l''ll'})
  \nonumber\\ && \quad
  + \sum_{ll'}(\bar v_{k_1lk_3l'}C_{k_4l'k_2l} + \bar v_{k_4lk_2l'}C_{k_1l'k_3l})
  \nonumber\\ && \quad
  - \frac{1}{2}\sum_{ll'}(\bar v_{k_1k_4ll'}C_{ll'k_2k_3} + \bar v_{ll'k_2k_3}C_{k_1k_4ll'})\Big]N^{-1}_{k_3k_4},
\label{K-0}
\end{eqnarray}
where
\begin{eqnarray}
  C_{k_1k_2k_3k_4}&=& \langle 0| c^\dag_{k_3}c^\dag_{k_4}c_{k_2}c_{k_1}|0\rangle
  \nonumber\\ &&
  - n_{k_1}n_{k_2}(\delta_{k_1k_3}\delta_{k_2k_4} - (k_3\leftrightarrow k_4))
  \label{C}
  \end{eqnarray}
is the fully correlated part (i.e. the cumulant) of the two-body density matrix.
We see that $K^0$ is divided into four parts: the first term on the right-hand side is the usual RPA  antisymmetrized interaction term. We should realize that in this first term the norm factor on the right of the interaction has been divided out [see Eq. (\ref{kernel2-static})] and that, contrary to standard RPA, the occupation factors are in principle not the HF ones but the correlated ones. Neglecting all the terms involving $C$ in Eq. (\ref{K-0}) but keeping correlations in the occupancies leads to the so-called renormalized RPA (r-RPA) briefly explained further in the Appendix. The next two terms are the one-body self-energy contributions (either the hole or the particle is not connected to the interaction). The remaining two-body correlation terms connect particles and holes. They can be qualified as screening terms and we want to investigate them further. The screening terms can be divided into two groups: the first two terms correspond to an exchange of \textit{p-h} fluctuations between the particle and hole and are, therefore, responsible for the screening of the long-range part of the interaction. 
This can be seen from the ordering of the indices $k_{1}$ and $k_3$ in the matrix element of the interaction. Clearly a creator and a destructor are correlated.
The second two terms correspond to an exchange of \textit{p-p}/\textit{h-h} fluctuations, that is they sum \textit{p-p}/\textit{h-h} ladder diagrams. They take care of the short-range correlations. 
Let us mention that neglecting the dynamic kernel, a self-consistent scheme for the two-body correlation function can be established, since it is given by integrating $R(\omega)$ over the frequency in the upper/lower half complex plane. This self-consistent scheme is referred to as SCRPA. It has the nice quality that all desirable properties of standard RPA, such as the fulfillment of the sum rule and conservation laws are maintained. This is explicitly shown in Ref.~\cite{Sch-epja}. In the past, it has produced encouraging results for several non-trivial model cases \cite{NPA512, Hubbard, 3-Lipkin, Tohy1}. Let us mention that Eq. (\ref{K-0}) has been given earlier \cite{NPA628} and that it has recently also been derived by Chatterjee and Pernal \cite{Pernal} for applications in chemical physics including, however, also diagonal configurations. 

In order to establish a connection with the static screened interaction $W$ of the $GW$+BSE approach, we consider in more detail the \textit{p-h} fluctuation terms. As an example, let us consider the fourth term on the right-hand side of Eq. (\ref{K-0}) and evaluate it first to second order in the interaction. Since $C$ is at least of first order, we will elaborate this and get the corresponding $K^0$ to second order. First let us give the relation between $C$ and the linear-response function $R$
\begin{equation}
  C_{k_4l'k_2l} = \langle 0|c^\dagger_lc_{l'}c^\dagger_{k_2}c_{k_4}|0\rangle - \bar n_{k_2}n_{k_4}\delta_{l'k_2}\delta_{lk_4} -n_ln_{k_2}\delta_{k_2k_4}\delta_{ll'},
  \label{C-R}
\end{equation}
where $\bar n_{k}=1 -n_{k}$ and
\begin{equation}
  \langle 0|c^\dagger_l c_{l'}c^\dagger_{k_2}c_{k_4}|0\rangle = i\lim_{t-t' \rightarrow 0^+} R_{l'lk_2k_4}(t-t') + n_ln_{k_2}\delta_{k_2k_4}\delta_{ll'}.
  \label{C-R2}
  \end{equation}
The reader may wonder why there is the last term on the right-hand side of above Eq. (\ref{C-R2}). The point is that since $R$ is the solution of the BSE-like equation, it does not contain such disconnected terms where the time $t$ does not communicate with time $t'$, see the definition of $R$ in Eq. (\ref{response}). This is also easily seen in solving, e.g., Eq.~(\ref{BSE}) to lowest order, that is without the kernel $K$ and using the HF form of norm $N$, which leads to Eq.~(\ref{R-0}) below. However, on the left-hand side, in the expectation value of the two-body-density-matrix operators, such terms are contained and, therefore, we have to add them on the right-hand side. A good way to see this is to evaluate Eq. (\ref{C-R}) in the HF approximation where $C=0$ by definition. Then the right-hand side must also be zero which is only the case if the extra term is added.
Let us now expand the response function in Eq. (\ref{C-R2}) up to first order
\begin{eqnarray}
  R_{l'lk_2k_4}(t-t') &=& R^{0}_{k_2k_4}(t-t')\delta_{l'k_2}\delta_{lk_4} +
\nonumber\\
&&\int dt_1 R^{0}_{l'l}(t-t_1)\bar v_{l'k_4lk_2}R^{0}_{k_2k_4}(t_1-t'),
\nonumber\\
  \label{v-1}
\end{eqnarray}
where $R^0$ is the non-interacting HF linear-response function
\begin{eqnarray}
  R^{0}_{k_2k_4}(t-t') &=& -i [\theta(t-t')\bar n^0_{k_2}n^0_{k_4} +
    \theta(t'-t)n^0_{k_2}\bar n^0_{k_4}]
\nonumber\\
&&\;\;\;\;\;\; \times e^{-i(\tilde e_{k_2}-\tilde e_{k_4})(t-t')}.
  \label{R-0}
\end{eqnarray}
Inserting Eq. (\ref{v-1}) into Eq. (\ref{C-R2}) and then Eq. (\ref{C-R2}) into Eq. (\ref{C-R}), and using $N^0_{k_1k_2} =N^{0^{-1}}_{k_1k_2}$, one obtains for the fourth term in the $K^0$ kernel
\begin{eqnarray}
K^{0,4}_{p_1h_2h_3p_4} =
    \sum_{ll'}\bar v_{p_1lh_3l'} C_{p_4l'h_2l}N^0_{h_3p_4}
  \nonumber\\ 
    \simeq \sum_{ll'}\bar v_{p_1lh_3l'}\frac{\bar n^0_{p_4}\bar n^0_{l'}n^0_{h_2}n^0_l}{\tilde \epsilon_{p_4} +\tilde \epsilon_{l'}-\tilde \epsilon_{h_2}-\tilde \epsilon_{l}}\bar v_{l'p_4lh_2}. \;\;\;\;
  \label{2nd-1}
\end{eqnarray}
Please note that the lowest-order term in Eq. (\ref{v-1}) is cancelled by the second term on the right-hand side of Eq. (\ref{C-R2}).
The expressions (\ref{2nd-1}) and (\ref{2nd-2}) (see below for the latter) are the only ones which contribute to $K^0$ at second order with a \textit{p-h} bubble exchange. We see this from the fact that the index $k_2$ in Eq. (\ref{2nd-1}) is a hole, then $k_1$ must be a particle and, since $k_4$ is a particle, $k_3$ must be a hole because our convention is that the index couple $k_1k_2$ or $k_3k_4$ can only be \textit{p}-\textit{h} or \textit{h}-\textit{p}. As we see, the term in Eq. (\ref{2nd-1}) contributes to the $B$ matrix in Eq. (\ref{A-B}). In analogy, we obtain for the fifth term in Eq. (\ref{K-0})
\begin{eqnarray}
K^{0,5}_{p_1h_2h_3p_4} = 
    \sum_{ll'}\bar v_{p_4lh_2l'} C_{p_1l'h_3l}N^0_{h_3p_4}
  \nonumber\\ 
    \simeq  \sum_{ll'}\bar v_{p_4lh_2l'}\frac{\bar n^0_{p_1}\bar n^0_{l'}n^0_{h_3}n^0_{l}}{\tilde \epsilon_{p_1} + \tilde \epsilon_{l'}-\tilde \epsilon_{h_3}-\tilde \epsilon_{l}}\bar v_{l'p_1 l h_3}.
  \label{2nd-2}
\end{eqnarray}
Again this term only contributes to the $B$ matrix of Eq.~(\ref{A-B}). Both terms correspond to the first two terms in Eq. (34) of Ref. \cite{Julien}. If we treat the last two (\textit{p-p}/\textit{h-h}) terms of our Eq. (\ref{K-0}) in the same way as the \textit{p-h} terms, we also reproduce the other two terms in Eq. (34) of Ref. \cite{Julien}
\begin{eqnarray}
(K^{0,6}+K^{0,7})_{p_1h_2h_3p_4} = 
\nonumber\\
  \frac{1}{2}\sum_{ll'}\bigg [\bar v_{p_1p_4ll'}\frac{n^0_{h_3}n^0_{h_2}\bar n^0_{l}\bar n^0_{l'}}{\tilde \epsilon_{h_3}
      + \tilde \epsilon_{h_2} - \tilde \epsilon_{l}-\tilde \epsilon_{l'}}\bar v_{ll'h_2h_3}\nonumber\\
    - \bar v_{ll'h_2h_3}\frac{ n^0_{l} n^0_{l'}\bar n^0_{p_1}\bar n^0_{p_4}}{\tilde \epsilon_{p_1} + \tilde \epsilon_{p_4} - \tilde \epsilon_l -\tilde \epsilon_{l'}}\bar v_{p_1p_4ll'}\bigg ].
  \label{2nd-3}
  \end{eqnarray}
As before, these terms only contribute to the $B$ matrix of Eq.~(\ref{A-B}).

From Eq. (\ref{K-0}), it is clear that in Eq. (\ref{v-1}) we can replace $R^0_{ll'}\delta_{ll_1}\delta_{l'l_3}$ by the full linear-response function $R_{ll'l_1l_3}$ what leads to a better approximation where the exchange \textit{p-h} bubble $ll'$ is replaced, e.g., by the RPA or even higher approximations. In general we have for $R$ in frequency space
\begin{eqnarray}
&&
R_{k_1k_2k_1'k_2'}(\omega) \equiv R^>_{k_1k_2k_1'k_2'}(\omega) - R^<_{k_1k_2k_1'k_2'}(\omega)=\nonumber\\
&&
\sum_{\nu}\frac{\langle 0|c^{\dag}_{k_2}c_{k_1}|\nu \rangle \langle \nu|c^{\dag}_{k_1'}c_{k_2'}|0\rangle }{\omega -\Omega_{\nu} + i\eta}-
\frac{\langle 0|c^{\dag}_{k_1'}c_{k_2'}|\nu \rangle \langle \nu |c^{\dag}_{k_2}c_{k_1}|0\rangle }{\omega +\Omega_{\nu} - i\eta},
\nonumber\\
\label{spectral-R}
\end{eqnarray}
where $\eta\to 0^+$. Actually this can be done also in Eq. (\ref{2nd-3}) where one can resum the pp ladders taking care of the short-range correlations. We will not further dwell on those extensions of our formalism for the moment.

Let us now consider the self-energy corrections in  Eq. (\ref{K-0}). For instance, let us extract a further interaction. For example, we obtain (indicating the time variables as subscripts for compactness) 
\begin{eqnarray}
    C_{k_4l''ll'} \simeq  -i \lim_{t'-t \rightarrow 0^+}\int dt_1 \; \theta(t_1 - t)n^0_{k_4} \;\;\;\;\;
  \nonumber\\  \qquad
  e^{-i\tilde \epsilon_{k_4}(t-t_1)}\langle 0| \mathrm{T} \{ j_{k_4}(t_1)(c^\dag_lc^\dag_{l'}c_{l''})_{t'} \} |0\rangle,
  \label{C-eom}
  \end{eqnarray}
and an analogous expression for the renormalization of the particle line.
Evaluating, as before, $C$ to first order, one obtains for the second and third term
\begin{eqnarray}
  &&K^{0,2+3}_{p_1h_2p_3h_4}=\nonumber\\
  &&\frac{1}{2}\sum_{ll'l''}\bigg [\delta_{h_2h_4} \bar v_{p_1ll'l''}\frac{n^0_{l'}n^0_{l''}\bar n^0_l\bar n^0_{p_3}}{\tilde \epsilon_{p_3}+\tilde \epsilon_l - \tilde \epsilon_{l'} - \tilde \epsilon_{l''}}\bar v_{l''l'lp_3}\nonumber\\
      &-&\delta_{p_1p_3}\bar v_{h_2l''l'l}\frac{\bar n^0_l\bar n^0_{l'}n^0_{l''}n^0_{h_4}}{\tilde \epsilon_{h_4}+\tilde \epsilon_{l''} - \tilde \epsilon_{l'} - \tilde \epsilon_{l}}\bar v_{ll'l''h_4}\bigg ].
\end{eqnarray}
We realize that this expression contributes only to the $A$ matrix and that
the second-order contribution to $K^0$ is now complete.

Of course, as in the case of the screening terms, we also can sum the \textit{p-h} bubbles to a full linear-response function. For this, we should factorize the three-body propagator in Eq. (\ref{C-eom}) into a product of a response function and a one-body propagator
\begin{eqnarray}
  &&
    \langle 0| \mathrm{T} \{ (c^\dag_{a_2}c_{a_4}c_{a_3})_{t_1}(c^\dag_lc^\dag_{l'}c_{l''})_{t'} \} |0\rangle \simeq
  \nonumber\\ && \qquad
    \Big[\langle 0| \mathrm{T} \{ (c^\dag_{a_2}c_{a_4})_{t_1}(c^\dag_{l'}c_{l''})_{t'} \} |0\rangle
   \langle 0| \mathrm{T} \{ c_{a_3}(t_1)c^\dag_l(t') \} |0\rangle 
  \nonumber\\ && \qquad
   - (a_3 \leftrightarrow a_4)\Big]
  - [l \leftrightarrow l'].
  \label{3bodyprogapprox}
\label{factor}
\end{eqnarray}
As indicated, there are four different ways to do this factorization and, thus, we multiply the final expression with this factor to obtain
\begin{eqnarray}
  &&
    -\frac{1}{2}\sum_{ll'l''} \delta_{k_1 k_3}\bar v_{ll'k_2l''}C_{k_4l''ll'} \simeq -\delta_{k_1 k_3}n^0_{k_4}
  \nonumber\\ &&
  \times\sum _{a_2a_3a_4}\sum_{ll'l''} \bar v_{k_4a_2a_3a_4}R^<_{a_2a_4l''l'}(\omega = \tilde \epsilon_{k_4}-\tilde \epsilon_{a_3}) \bar v_{ll'k_2l''}.
  \nonumber\\
  \label{R-stat-self}
\end{eqnarray}
Proceeding with the second self-energy correction in the same way, one obtains  an analogous expression.

We will discuss the relation with the static form of $W$ below in Sec. IV.
The well-known problem coming from the approximation of Eq.~(\ref{3bodyprogapprox}) is that if the response function is replaced by its uncorrelated \textit{p-h} response in Eq.~(\ref{R-stat-self}), one does not recover the correct second-order expression of the kernel \cite{Pawel}. The result is by a factor of two too large and, therefore, one usually subtracts the second-order contribution in order to obtain the correct lowest-order contribution of $\Sigma$ to the kernel and also of the ensuing RPA correlation energy \cite{Pawel}. In principle this subtraction procedure is not completely correct, since the corresponding imaginary part of the self-energy has no definite sign. How this can be fixed is explained in Refs. \cite{Pawel, Adachi}. We will not dwell on this here and ignore this subtlety in the remainder of the paper, supposing that the uncorrelated terms are small, and we will concentrate on the comparison with approximations given in the literature \cite{Pina,Pina2,Julien} where this problem is also not addressed.

In our approach, the one-body self-energy corrections appear directly in the purely static part of the integral kernel. So the self-energy corrections are treated consistently with the screening terms. This may be important because it is known that often there are significant cancellations between both contributions. Since screening and self-energy corrections in the purely static part $K^0$ of the kernel involve again the response function $R$, as explained above,  a self-consistent cycle can be established.

\subsection{The dynamic part of the one-frequency kernel $K$}

Let us now discuss the {\it time-dependent (dynamic)} part $K^\mathrm{dyn}$ of the interaction kernel in Eq.~(\ref{kernel2}) which is a \textit{p-h} irreducible four-body propagator of the \textit{2p-2h} type. It is straightforward to evaluate it to lowest order. Since the dynamic kernel is already explicitly of second order in the interaction, it is sufficient to evaluate the \textit{2p-2h} propagator to lowest (HF) order. Using $J_{k_1k_2} = c_{k_2}^\dag j_{k_1}- j^\dag_{k_2} c_{k_1}$ and dropping for now, for simplicity, the factor $N_{k_1' k_2'}^{{0^{-1}}}$ which just gives a minus sign for the contribution to the $B$ matrix, the full expression of the second-order dynamic kernel is then
\begin{eqnarray}
    K^{{\rm dyn}, (2)}_{k_1k_2k'_1k'_2}(t-t') &=& 
     -i\langle 0 | \mathrm{T} \{ 
    (c^\dag_{k_2}j_{k_1})_t (j^\dag_{k_1'}c_{k'_2})_{t'} \}| 0 \rangle_0
     \nonumber\\ &&
    -i\langle 0 | \mathrm{T} \{ 
    (j^\dag_{k_2}c_{k_1})_t (c^\dag_{k'_1}j_{k'_2})_{t'} \}| 0 \rangle_0
     \nonumber\\ &&
   +i\langle 0 | \mathrm{T} \{ 
    (c^\dag_{k_2}j_{k_1})_t (c^\dag_{k'_1}j_{k'_2})_{t'} \}| 0 \rangle_0
    \nonumber\\ &&
    +i\langle 0 | \mathrm{T} \{ 
    (j^\dag_{k_2}c_{k_1})_t (j^\dag_{k_1'}c_{k'_2})_{t'} \}| 0 \rangle_0,
    \nonumber\\
  \label{K-dyn-0}
  \end{eqnarray}
where the subscript ``0'' indicates that this term is evaluated to lowest order.
The first two terms are self-energy corrections recognizable by the index pair $k_1, k'_1$ or $k_2, k'_2$ whereas the other two terms have ``mixed'' indices. These expressions describe the decay of a \textit{p-h} mode into uncorrelated (incoherent) \textit{2p-2h} states. The four terms have different meanings. The two terms with either $j j^\dag$ or $j^\dag j$ describe, as just mentioned, dynamic self-energy corrections to the particle and the hole states, respectively. The other two terms with $jj$ or $j^\dag j^\dag$ describe a \textit{p-h} exchange between the particle and the hole. Such incoherent processes have already been considered a long time ago by Landau \cite{Landau} in his study of the damping of zero sound in a Fermi liquid. A detailed study of this is given in Ref. \cite{Adachi}. 

To obtain the spectral representation of  $K^{{\rm dyn}, (2)}$, we just need to consider the generic propagator
\begin{eqnarray}
  &&-i\langle 0| \mathrm{T} \{ (c^\dag_{k_4}c^\dag_{k_3}c_{k_1}c_{k_2})_t(c^\dag_{k'_2}c^\dag_{k'_1}c_{k'_3}c_{k'_4})_{t'} \}| 0 \rangle_0, 
  \label{2p-2h}
\end{eqnarray}
calculate its Fourier transform
\begin{eqnarray}
  &\bigg [&\frac{n^0_{k_4}n^0_{k_3}\bar n^0_{k_1}\bar n^0_{k_2}}{\omega - \tilde \epsilon_{k_1}-\tilde \epsilon_{k_2}+\tilde \epsilon_{k_4} +\tilde \epsilon_{k_3} +i\eta}\nonumber\\
    &-&   \frac{\bar n^0_{k_4}\bar n^0_{k_3} n^0_{k_1} n^0_{k_2}}{\omega - \tilde \epsilon_{k_1}-\tilde \epsilon_{k_2}+\tilde \epsilon_{k_4} +\tilde \epsilon_{k_3} -i\eta}\bigg ]\delta_{k_1k_2k'_1k'_2}\delta_{k_3k_4k'_3k'_4},\nonumber\\
  &~&
  \label{2p-2hfreq}
\end{eqnarray}
with $\delta_{k_1k_2k'_1k'_2} = \delta_{k_1k'_1}\delta_{k_2k'_2}-\delta_{k_1k'_2}\delta_{k_2k'_1}$, and use this in Eq. (\ref{K-dyn-0}). The obtained expression is well known in the nuclear physics literature \cite{Wambach}. More recently such expressions have been derived by Rebolini and Toulouse \cite{Julien}, but starting from the Hedin equations and without including the self-energy corrections.

Instead of approximating the \textit{2p-2h} propagator by its uncorrelated expression, we can include higher-order effects. For example one can factorize it into a product of a response function and an uncorrelated \textit{p-h} propagator. Or, one can factorize it into a product of two linear-response functions.
The choice will depend on the physical situation.
Such approximations have been considered in Ref. \cite{Schuck}. As in the case of the self-energy of the one-body Dyson equation, those factorizations do not give, however, the correct lowest-order limit of the kernel. If important, one has to correct for it. How this can be done consistently is explained, as already mentioned, in Refs. \cite{Pawel, Adachi}.
Let us give explicit expressions for the spectral representation of $K^{\rm dyn}$ for the case where we approximate the \textit{2p-2h} propagator into a product of a linear-response function times an uncorrelated \textit{p-h} propagator.
Typically, one will evaluate the response function  with the RPA method. It is then easy to get the spectral form of $K^{\rm dyn}$ (skipping now the \textit{p-p}/\textit{h-h} contributions)
\begin{eqnarray}
  &K&^{{\rm dyn}}_{k_1k_2k'_1k'_2}(\omega)=
  \sum_{\nu}\bigg \{ \sum_{l_2l_3l_4l'_2l'_3l'_4}\nonumber\\
  &\bigg (&\delta_{k_2,k'_2}\bar v_{k_1l_2l_3l_4}\frac{ \langle 0|c^\dag_{l_2}c_{l_4}|\nu\rangle \langle \nu|c^\dag_{l'_4}c_{l'_2}|0\rangle}{\omega -(\tilde \epsilon_{l_3} - \tilde \epsilon_{k_2} + \Omega_{\nu})
    + i\eta}\bar v_{l'_4l'_3l_3l'_2k'_1}\nonumber\\
 &+& \delta_{k_1,k'_1}\bar v_{k_2l_2l_3l_4}\frac{ \langle 0|c^\dag_{l_4}c_{l_2}|\nu\rangle \langle \nu|c^\dag_{l'_2}c_{l'_4}|0\rangle}{\omega -(\tilde \epsilon_{k_1} - \tilde \epsilon_{l_3} + \Omega_{\nu}) + i\eta}\bar v_{l'_4l'_3l_3l'_2k'_2}\bigg )
  \nonumber\\
  &+&   \sum_{l_2l_4l'_2l'_4}\bigg (\bar v_{k_1l_2k'_1l_4}\frac{ \langle 0|c^\dag_{l_2}c_{l_4}|\nu\rangle \langle \nu|c^\dag_{l'_2}c_{l'_4}|0\rangle}{\omega -(\tilde \epsilon_{k'_1} - \tilde \epsilon_{k_2} + \Omega_{\nu}) + i\eta}\bar v_{k'_2l'_2k_2l'_4}\nonumber\\
  &+&\bar v_{k_2l_2k'_2l_4}\frac{ \langle 0|c^\dag_{l_4}c_{l_2}|\nu\rangle \langle \nu|c^\dag_{l'_4}c_{l'_2}|0\rangle}{\omega -(\tilde \epsilon_{k_1} - \tilde \epsilon_{k'_2} + \Omega_{\nu}) + i\eta}\bar v_{l'_4k_1l'_2k'_1}\bigg )\bigg \}.
  \label{K-dyn-2}
  \end{eqnarray}
We note that the first two terms on the right-hand side of the above equation correspond again to self-energy corrections, whereas the last two terms are contributions where \textit{p-h} modes are exchanged between the particle and the hole. We also realize that these exchange contributions correspond to Eq. (31) of Ref. \cite{Pina}. The ``backward going'' terms do not contribute as easily realized. If we consider the static limit ($\omega=0$), they should be considered together with Eqs. (\ref{2nd-1}) and (\ref{2nd-2}). This shall be discussed in more detail in the next section.

Before doing so, it may be worth showing how to include further \textit{p-h} correlations in summing up the free \textit{p-h} propagators, contained in Eq. (\ref{K-dyn-2}), to extra RPA modes. This is most easily done by factorizing the \textit{2p-2h} propagator into a fully antisymmetrized product of two \textit{p-h} response functions
\begin{eqnarray}
  -&i&\langle 0| \mathrm{T} \{ (c^\dag_{k_4}c^\dag_{k_3}c_{k_1}c_{k_2})_t(c^\dag_{k'_2}c^\dag_{k'_1}c_{k'_3}c_{k'_4})_{t'} \} | 0 \rangle \simeq \nonumber\\
  &i&\Big [\big\{[R_{k_2k_4k'_2k'_4}(t-t') R_{k_1k_3k'_1k'_3}(t-t')
      -(k'_3 \leftrightarrow k'_4)]\nonumber\\
    &-&[k'_1 \leftrightarrow k'_2]\big\} - \{k_1 \leftrightarrow k_2\}\Big ] - \Big [ k_3 \leftrightarrow k_4 \Big ], 
  \label{double-RPA}
\end{eqnarray}
where
\begin{eqnarray}
  R_{k_2k_4k'_2k'_4}(t-t') &=& \sum_{\nu}\bigg [ \langle 0|c^\dag_{k_4}c_{k_2}|\nu\rangle \langle \nu|c^\dag_{k'_2}c_{k'_4}|0\rangle e^{-i\Omega_{\nu}(t-t')}\nonumber\\
    &+&\langle 0|c^\dag_{k'_2}c_{k'_4}|\nu\rangle \langle \nu|c^\dag_{k_4}c_{k_2}|0\rangle e^{i\Omega_{\nu}(t-t')} \bigg ]
  \label{spectral-t}
\end{eqnarray}
is the Fourier transform into time space of Eq. (\ref{spectral-R}).
The Fourier transform of Eq. (\ref{double-RPA}) into frequency space is then easily performed with Eq. (\ref{spectral-t})
\begin{eqnarray}
  && \sum_{\nu\nu'} \bigg [ \frac{\langle 0|c^\dag_{k_4}c_{k_2}|\nu\rangle \langle \nu|c^\dag_{k'_2}c_{k'_4}|0\rangle \langle 0|c^\dag_{k_3}c_{k_1}|\nu'\rangle \langle \nu'|c^\dag_{k'_1}c_{k'_3}|0\rangle}{\omega - \Omega_{\nu}-\Omega_{\nu'} +i\eta}
  \nonumber\\ &&
    - \frac{\langle 0|c^\dag_{k'_2}c_{k'_4}|\nu\rangle \langle \nu|c^\dag_{k_4}c_{k_2}|0\rangle \langle 0|c^\dag_{k'_1}c_{k'_3}|\nu'\rangle \langle \nu'|c^\dag_{k_3}c_{k_1}|0\rangle  }{ \omega + \Omega_{\nu}+\Omega_{\nu'} -i\eta }\bigg ]
  \nonumber\\ &&
  + \quad {\rm exchange~ terms},
  \label{F-d-RPA}
\end{eqnarray}
where ``exchange terms'' means that all exchange terms present in Eq. (\ref{double-RPA}) should be included also here.
Inserting Eq. (\ref{F-d-RPA}) into Eq. (\ref{kernel2}) yields an expression equivalent to Eq. (23) of Ref. \cite{Pina2} (see also Ref. \cite{Schuck}). Notably only the first term with $+i\eta$ will survive, that is, it enters only the $A$ matrix, as also pointed out in Ref. \cite{Pina2}. Since it is fully antisymmetric between the two-particle states and two-hole states in entrance and exit channels, the approximation gives a conserving approximation for the response function \cite{KB,Sch-epja}.

\section{Comparison with $GW$+BSE}
\label{comparison}

Let us now consider similarities and differences of the present approach to the response function and the $GW$+BSE scheme as commonly used in condensed-matter and chemical physics.

A first point consists in the fact that in the present formalism all Coulomb matrix elements are antisymmetrized [see Eq. (\ref{as-v})] whereas in the $GW$+BSE scheme all exchange matrix elements are usually absent besides the one contained in the first order of the screening term. This also concerns the $W$ used within the RPA in condensed-matter physics: only the bubble diagrams are resummed, as it was done in the original work of Bohm and Pines \cite{Pines}. Including then the static limit of $K^{\rm dyn}$ (i.e. at $\omega = 0$) in Eq. (\ref{K-dyn-2}) to $K^0$ of Eq. (\ref{K-0}) yields an expression very similar to the ``excitonic'' Hamiltonian $H^{2p,exc}$ in Eqs. (16) and (21) of Ref. \cite{Pina}. However, there are also substantial differences and, for a detailed comparison, let us give our full static expression here (summing the \textit{p-h} bubble exchange to a full response function and skipping the self-energy and \textit{p-p}/\textit{h-h} contributions for easier comparison)
\begin{eqnarray}
  &K&^{\rm stat}_{k_1k_2k'_1k'_2} = v_{k_1k_2k'_1k'_2} - v_{k_1k_2k'_2k'_1}\nonumber\\
  &-& \sum_{l_1l'_1}\sum_{l_2l'_2}\bigg (\nonumber\\
  &\bar n&\!^0_{k_1}n^0_{k'_1}\bar v_{k_1l_1k'_1l'_1}\sum_{\nu}\frac{ \langle 0|c^\dag_{l'_1}c_{l_1}|\nu \rangle \langle \nu c^\dag_{l_2}c_{l'_2}|0 \rangle}{\tilde \epsilon_{k_2} - \tilde \epsilon_{k'_2} + \Omega_{\nu} -i\eta}\bar v_{l'_2k_2l_2k'_2}\bar n^0_{k'_2}n^0_{k_2}\nonumber\\
&+& \nonumber\\
  &\bar n&\!^0_{k_2}n^0_{k'_2}\bar v_{k_2l_1k'_2l'_1}\sum_{\nu}\frac{ \langle 0|c^\dag_{l'_1}c_{l_1}|\nu \rangle \langle \nu c^\dag_{l_2}c_{l'_2}|0 \rangle}{\tilde \epsilon_{k_2} - \tilde \epsilon_{k'_2} + \Omega_{\nu} -i\eta}\bar v_{l'_2k_1l_2k'_1}\bar n^0_{k'_1}n^0_{k_1}\bigg )\nonumber\\
  &-&
  \sum_{l_2l_4l'_2l'_4}\bigg (\nonumber\\
  &\bar n&\!^0_{k_1}\bar n^0_{k'_1}\bar v_{k_1l_2k'_1l_4}\frac{ \langle 0|c^\dag_{l_2}c_{l_4}|\nu\rangle \langle \nu|c^\dag_{l'_2}c_{l'_4}|0\rangle}{\tilde \epsilon_{k'_1} - \tilde \epsilon_{k_2} + \Omega_{\nu} + i\eta}\bar v_{k'_2l'_2k_2l'_4}n^0_{k_2}n^0_{k'_2}\nonumber\\
&+&\nonumber\\
  &n&\!^0_{k_2}n^0_{k'_2} \bar v_{k_2l_2k'_2l_4}\frac{ \langle 0|c^\dag_{l_4}c_{l_2}|\nu\rangle \langle \nu|c^\dag_{l'_4}c_{l'_2}|0\rangle}{\tilde \epsilon_{k_1} - \tilde \epsilon_{k'_2} + \Omega_{\nu} + i\eta}\bar v_{l'_4k_1l'_2k'_1}\bar n^0_{k_1}\bar n^0_{k'_1}\bigg ).\nonumber\\
  &~&
\label{K-stat}
  \end{eqnarray}
We see that the first two terms belong to the $B$ matrix and the last two terms to the $A$ matrix of Eq. (\ref{A-B}). In the $GW$+BSE scheme all antisymmetrized matrix elements $\bar v_{k_1k_2k_3k_4}$ in Eq.~(\ref{K-stat}) are replaced by only the direct term $v_{k_1k_2k_3k_4}$. In addition, in the denominators the differences of orbital energies are absent, so that only the RPA roots $\Omega_{\nu}$ remain, which corresponds to the static $W(0)$ of the $GW$+BSE kernel (see, e.g., Ref. \cite{Pina}). It is difficult to judge the combined effect of the two differences of $GW$+BSE with respect to the above expression in Eq. (\ref{K-stat}). The extra orbital energies in the denominators in our expressions have, however, certainly a reduction effect. A detailed numerical evaluation is out of the scope of the present work but shall eventually be presented in the future. It seems to us that the appearance of the orbital energies in the denominators has its justification. In the work of Romaniello \textit{et al.} \cite{Pina} they also appear as an extra static contribution from their $\widetilde W$ expression in Eq. (27) in Ref. \cite{Pina} in putting therein $\omega_{\lambda}=0$. It is clear that, although the terms in Eq. (\ref{K-stat}) are instantaneous, there can never be an {\it exact} equal time process when an RPA mode crosses between the particle and the hole lines. There is always an infinitesimal time difference allowing for the orbital energies to appear in the denominators of Eq. (\ref{K-stat}).

A further difference of our EOM approach is that the self-energy contributions appear directly in the kernel. It is possible to resum them separately, which would lead to dressed quasi-particles (and quasi-holes), quite similarly to the $GW$+BSE scheme. In this respect we do not see any significant difference between the two approaches.

In our scheme we obtain the same (approximate) dynamic contributions to the kernel as obtained by Sangalli \textit{et al.} \cite{Pina2} [see their Eqs. (22) and (23)]. They also contain the self-energy contributions. It is also clear that those dynamic contributions only renormalize the $A$ matrix and give no contribution to the $B$ matrix. On the other hand in Ref. \cite{Pina2}, the $B$ matrix is not renormalized, not containing the additional correlations which are summed up in Eq. (\ref{K-0}). In Ref. \cite{Julien}, the renormalization of the $B$ matrix is given only to lowest order.
Let us also point out that the so-called time-blocking approximation (TBA) \cite{Litvinova} invented recently in nuclear physics to derive a kernel depending only on one frequency certainly has a very close relation with the procedures employed in Refs. \cite{Julien} and \cite{Pina, Pina2}. It may be relevant to realize that the first two terms in Eq. (\ref{K-stat}) which derive from Eq. (\ref{K-0}) and renormalize the $B$ matrix are an approximation to Eq. (\ref{K-0}) [see the Appendix]. As we will see in the next section in some two-body problems it may be important to keep the full expression of Eq. (\ref{K-0}).

\section{Illustration on the Hubbard molecule}
\label{Hubbardmol}

The Hubbard model describes electrons on a lattice with the Coulomb interaction replaced by an on-site constant $U$. The well-known Hamiltonian is given by
\begin{equation}
  H=-t\sum_{<ij>\sigma}c^\dag_{i\sigma}c_{j\sigma} + U\sum_i\hat n_{i\uparrow}\hat n_{i\downarrow},
  \label{Hubb-1}
\end{equation}
where $c^\dag_{i\sigma}$ and $c_{i\sigma}$ are the electron creation and destruction operators at site $i$ with spin projection $\sigma$ and the $\hat n_{i\sigma} = c^\dag_{i\sigma}c_{i\sigma}$ are the number operators for electrons at site $i$ with spin projection $\sigma$. As usual $t$ is the nearest-neighbor hopping integral. For demonstration purposes, in this work, we will limit ourselves to the simplest non-trivial case which is the one of two sites ($N_\text{s}=2$) with two electrons, the so-called Hubbard molecule. As  the problem has already been solved exactly with the SCRPA method \cite{Hubbard} derived from Rowe's \cite{Rowe} EOM, we only will outline the basic principle here using, however, the present approach. It is advantageous to write the Hamiltonian in momentum space (we consider periodic boundary conditions)
  \begin{equation}
    H= \sum_{{\bf k},\sigma}(\epsilon_k - \mu)\hat n_{{\bf k},\sigma} + \frac{U}{2N_\text{s}}\sum_{{\bf k},{\bf p},{\bf q},\sigma} c^\dag_{{\bf k},\sigma}c_{{\bf k}+{\bf q},\sigma}c^\dag_{{\bf p},-\sigma}c_{{\bf p}-{\bf q},-\sigma},
    \label{Hubb-2}
  \end{equation}
  where $\hat n_{{\bf k},\sigma} = c^\dag_{{\bf k},\sigma}c_{{\bf k},\sigma}$ is the occupation number operator of the momentum-spin mode $({\bf k},\sigma )$ and $\epsilon_{k}= - 2t\cos(k)$ are the one-body energies with the lattice spacing set to unity. Because of having only two electrons and the periodic boundary conditions, the only allowed momenta are $k_1 = 0$ and $k_2 = -\pi$. Accordingly, we only have two types of \textit{p-h} operators: $J_{\sigma} = c^\dag_{k_1,\sigma}c_{k_2,\sigma}$ with $\sigma = \pm 1/2$. Let us introduce the ``charge'' and ``spin'' operators
\begin{equation}
J^{(\pm)} = J_{\uparrow} \pm J_{\downarrow},
\end{equation}
  and consider the charge and spin linear-response functions
  \begin{eqnarray}
 {R}^{(\pm)}(t-t') = \;\;\;\;\;\;\;\;\;\;\;\;\;\;\;\;\;\;\;\;\;\;\;\;\;\;\;\;\;\;\;\;\;\;\;\;\;\;\;\;\;\;\;\;\;
\nonumber\\
 -i\langle 0|\text{T}
    \begin{pmatrix}J^{(\pm)}(t)J^{{(\pm)}^\dag}(t')&J^{(\pm)}(t)J^{{(\pm)}}(t')\\
      J^{{(\pm)}^\dag}(t)J^{{(\pm)}^\dag}(t')&J^{{(\pm)}^\dag}(t)J^{(\pm)}(t')\end{pmatrix}|0\rangle. \;\;
\label{R-Hubb}
\end{eqnarray}
 We therefore have to consider two $2\times 2$ matrix response functions. For this very simple example it so happens that the dynamic part $K^{\rm dyn}$ of the one-frequency kernel decouples from the purely static part $K^0$, and only $K^{0}$ contributes in the \textit{p-h}/\textit{h-p} space.  As seen from Eq. (\ref{K-0}), the purely static kernel $K^0$ only contains static two-body correlation functions. They can be calculated from integrating ${R}^{(\pm)}(\omega)$ over the frequency in the upper/lower half complex plane. Since additionally the occupation numbers can also be expressed via the static two-body correlation functions as (see Ref. \cite{Hubbard})
  \begin{equation}
    n_{p,\sigma} = \sum_h \langle 0 | J^\dag_{ph,\sigma} J_{ph,\sigma}|0 \rangle~,~~~~
    n_{h,\sigma} = \sum_p \langle 0|J^\dag_{ph,\sigma} J_{ph,\sigma}|0\rangle \,,
    \label{occs-Hubb}
        \end{equation}
we have a closed system of equations which can be solved. It turns out that the exact solution is obtained. This is explained in detail in Ref. \cite{Hubbard} starting, however, with the equivalent EOM for RPA operators and not with the Green functions and we will not repeat the whole procedure  here. 

The fact that the two-body problem is solved exactly by the SCRPA in the Hubbard model is also found in several other models, like the Lipkin model \cite{Tohy1} and the pairing model \cite{Tohy1,Hirsch}. However, it is not a general feature of SCRPA that it solves any two-body problem exactly. Generally there exist specific \textit{2p-2h} configurations which have to be taken into account when solving a two-body problem. It should also be pointed out that the two-body correlation functions in Eq. (\ref{K-0}) cannot be further approximated if the exact solution for, e.g., the Hubbard molecule shall be obtained. Already the forms in Eqs. (\ref{2nd-1}) and (\ref{2nd-2}) are approximations to Eq. (\ref{K-0}) even if the exchange bubble is resummed to a full linear-response function and it is likely that they will not maintain the exact solution. It is thus seen from this example that the present approach leads in a systematic way to manageable expressions which, if necessary, sum higher correlations than is the case with the $GW$+BSE approach.

Finally, a reduced version of SCRPA called r-RPA is presented in the Appendix.
In a separate paper \cite{Li18} this approximation has been applied to the case of a real system: the helium atom.
In that paper both the r-RPA and the $GW$+BSE solutions are compared to the exact Hylleraas solution evidencing their respective performances, similarities and differences, on a real system.

\section{Conclusions}
\label{conclusions}

The objective of this work was three-fold. First, we derived a formally exact one-frequency-only BSE-like equation for the linear-response function whose integral kernel only depends on the single frequency of the applied field. Explicit expressions of this kernel in terms of higher Green functions are presented. They lend themselves very naturally to physically motivated approximations. Second, in this way, known approximations of a single frequency kernel derived from Hedin's equations are straightforwardly recovered. It is shown that with our approach not only the second-order expressions for the static ($B$ matrix) and dynamic BSE kernel given in Ref. \cite{Julien} can be recovered but that these second-order terms can naturally be resummed to full linear-response functions. This has also been shown in Refs. \cite{Pina,Pina2} for the dynamic part but the renormalization of the static part ($B$ matrix), which is of the same order as the dynamic one, is missing there. Taking the static limit ($\omega=0$) of the dynamic part, we obtain a complete expression for the static limit of our kernel. Third, this then allows us to make a detailed comparison with the static limit of the kernel of the well known $GW$+BSE approach. Between both static approaches there exist, besides quite some similarities, also substantial differences which may be interesting to study further in future work with numerical examples. At the end of the paper, we also show that for the so-called Hubbard molecule the exact solution can be recovered from our approach. This is only possible with a consistent and fully resummed static kernel as presented here. Let us finally mention that the present formalism is very much related to the EOM introduced by Rowe and further elaborated in Ref.~\cite{Tohy1}.

\section{Acknowledgements}

P.S. is grateful to D. Delion, J. Dukelsky, M. Jemai, and M. Tohyama for a fruitful collaboration on the EOM method for correlation functions.
Useful comments on the manuscript from M. Holzmann, K. Pernal, and P. Romaniello are appreciated.

\appendix

\section*{Appendix: The renormalized RPA}
A very much simplified version of SCRPA consists in neglecting in $K^0$ all the $C$ correlation function terms. Then, one obtains for the one-frequency BSE-like equation
\begin{eqnarray}
R_{k_1k_2,k'_1k'_2}(\omega) &=& \tilde R^{0}_{k_1k_2}(\omega)\delta_{k_1k'_1}\delta_{k_2k'_2} 
\nonumber\\
&&+ \sum_{k_3k_4}\tilde R^{0}_{k_1k_2}(\omega)\bar v_{k_1k_4k_2k_3}R_{k_3k_4k'_1k'_2}(\omega),
\nonumber\\
\label{r-RPA}
\end{eqnarray}
with
\begin{equation}
    \tilde R^{0}_{k_1k_2}(\omega) = \frac{(1-n_{k_1})n_{k_2}}{\omega - (\tilde \epsilon_{k_1} - \tilde \epsilon_{k_2}) + i \eta}-\frac{n_{k_1}(1-n_{k_2})}{\omega - (\tilde \epsilon_{k_1} - \tilde \epsilon_{k_2}) - i \eta}.
    \label{R0}
\end{equation} 
We see that this renormalized RPA (r-RPA) equation is like the standard RPA besides the fact that the occupation numbers are the correlated ones and not the HF ones.
We thus have to give an expression for the $n_k$'s which couple back to the RPA.
Such an approximation for the occupation numbers $n_k$ has, e.g., been derived by Catara \textit{et al.} \cite{CatPicSamVan-PRB-96}. The expressions are given by
\begin{equation}
  n_h = 1- \langle 0 | c^\dag_hc_h | 0 \rangle ~~ =~~ \frac{1}{2}\sum_p\langle 0 | c^\dag_pc_hc^\dag_hc_p | 0 \rangle,
\end{equation}
and
\begin{equation}
  n_p = \langle 0 | c^\dag_pc_p | 0 \rangle ~~= ~~\frac{1}{2}\sum_h\langle 0 | c^\dag_pc_hc^\dag_hc_p | 0 \rangle,
  \end{equation}
where the two-body density matrix can directly be obtained from the linear-response function.
We, therefore, have established a minimal self-consistent system of equations where the occupation numbers are calculated from the response function.


%
\end{document}